\documentclass[fleqn,12pt,twoside]{article}
\usepackage{espcrc1}

\newcommand{\AmS}{{\protect\the\textfont2
  A\kern-.1667em\lower.5ex\hbox{M}\kern-.125emS}}
\newcommand{\beq}{\begin{equation}}
\newcommand{\eeq}{\end{equation}}
\newcommand{\rlong}{R_{long}}
\newcommand{\rout}{R_{out}}
\newcommand{\rside}{R_{side}}

\title{Theory at Quark Matter '02}

\author{Robert D. Pisarski
	\address{High Energy Theory, Bldg. 510A\\
	Brookhaven National Laboratory\\
	Upton, NY 11973-5000}
        \thanks{This research was supported by
	DOE grant DE-AC-02-98CH-10886.}}
       
\begin{document}

\maketitle

\begin{abstract}
I summarize theory at Quark Matter 2002,
stressing the continuing inability of a single model to describe
all notable features of the data from
$\sqrt{s}/A: 55 \rightarrow 200$~GeV.
\end{abstract}

\section{INTRODUCTION}

The collisions of large nuclei may give us insight into the nature
of QCD at high temperatures and/or densities.  In this talk I review
the status of theory for heavy ion collisions from Quark Matter 2002.
I concentrate on the the results for the largest nuclei, with
atomic number $A \approx 200$.

At the SPS at CERN, there are two notable results for $AA$ collisions,
for $\sqrt{s}/A: 5 \rightarrow 17$~GeV \cite{satz}:

{\it $J/\Psi$ suppression}: the number of $J/\Psi$ pairs is smaller
in the most central collisions, versus the extrapolation from
peripheral collisions, or from collisions with smaller $A$ \cite{ramello}.
The effect is most striking for the largest nuclei.  

{\it Excess dileptons below the $\rho$}:  the rate of $e^+e^-$
pairs exceeds that in conventional hadronic models
\cite{wessels,wambach}, although a broadened $\rho$-meson
explains the excess.  The effect is more prominent at lower, and
not high energies, suggesting a density dependent effect.  This
also supports interest in going to even lower energies, such
as at the proposed GSI collider.

At BNL, RHIC has run at energies of $\sqrt{s}/A = 55$~GeV (briefly),
at $130$~GeV during Run I, and at $200$~GeV during Run II.  Results
from Run I were first presented at Quark Matter 2001; those from
Run II, at this Quark Matter, 2002.  

There is one notable change
expected change between the SPS and RHIC.
At the SPS, the particle multiplicity in $AA$ collisions is
a single peak about zero rapidity.  By RHIC energies, a Central
Plateau was expected to open up, in which physics is (approximately)
boost invariant, independent of rapidity.
Away from the incident nucleons of the
fragmentation region,
the Central Plateau is where a system at nonzero temperature,
and almost zero quark density, might emerge.  Perhaps
even deconfined matter, as the Quark-Gluon Plasma.

RHIC experiments find that the Central Plateau is rather
narrow.  In all, particles are spread out over 
$\approx \pm 5$ units of rapidity.  The multiplicites for
identified particles are nearly constant over the central
$\approx \pm 1$ unit of rapidity \cite{bearden,ulrich,vanburen}.
However, the pion's average transverse momentum,
$p_t$, is only constant over $\pm .5$ units of rapidity
\cite{ulrich}.  But as with other variables,
there is much to gain by looking at all rapidities.

Even without (much) electromagnetic data, which proved so
interesting at the SPS, so far there are four notable features 
of the RHIC data:

{\it ``High''-$p_t$ suppression}: 
the number of particles with transverse momentum $p_t$ 
between $2-10$~GeV is suppressed, relative to that in $pp$,
times the number of binary collisions
\cite{kunde,miod,baier,muller,wang}.  The overall
suppression is by factors of $2-4$.
The suppression is now seen to be
approximately constant for these $p_t$ \cite{miod}.
This is opposite what happens at
the SPS, where high $p_t$ particles are not suppressed,
but enhanced by factors of $2-3$, through the Cronin effect.

{\it Elliptic flow}: is a measure of 
momentum anistropy in non-central collisions.  
Hydrodynamics predicts elliptic flow is linear in $p_t$ for pions.
This is seen up to $p_t \approx 1.5$~GeV, as is the hydrodynamic
behavior of protons.  For $p_t: 2 \rightarrow 6$~GeV, though,
the (total) elliptic flow is flat \cite{voloshin,huovinen},
which is not expected from hydrodynamics (or anything else).

{\it HBT radii}: pion interferometry gives a measure of the
spatial size(s) of the system.  Hydrodynamics predicts that a certain
ratio of two sizes, $R_{out}/R_{side}$, is greater than one,
and increases as $p_t$ does.  Instead,
experiment finds that $\rout/\rside$ decreases
with increasing $p_t$, and is about one by $p_t \approx 400$~MeV
\cite{huovinen,pratt}.  HBT radii indicate that hadronization
occurs as a type of ``blast'' wave \cite{pratt}.

{\it Jet absorption}: At these energies, 
jets are seen in $pp$ collisions, but an angular correlation finds that
in $AA$ collisions, the 
``backward'' jet is strongly suppressed \cite{hardtke}.  That
is, in $AA$ collisions there is stuff there which eats jets.

It is thus an extremely exciting time.  
For now, experiment is triumphant: especially after the run
at $\sqrt{s}/A = 200$~GeV, there is a striking agreement between
different experiments for many quantities of interest.  
The crucial advance has been precise measurements.  If quantities
are only measured to $\pm 30\%$, then a lot of theories can 
hide under the error bars.  With errors $\pm 5\%$, it becomes
possible to definitively rule (many) theories out.

The challenge
to theorists to synthesize {\it all} of these measurements into
a consistent framework.  
Common prejudices held before the RHIC data --- such as large
increases in multiplicity, and large sizes from a strongly
first order transition --- are extinct.
Certain models can explain some aspects
of the RHIC data, but at present, {\it no} model can
explain everything interesting.

My review of theory at Quark Matter 2002 is organized thematically,
emphasizing results of relevance to experiment.  
Consequently, I will offend by omission, neglecting all work on
color superconductivity and other interesting topics.  

\section{THE BEDROCK: THE LATTICE}

Our understanding of QCD at nonzero temperature rests upon
numerical simulations on the Lattice \cite{kanaya}.
While dynamical quarks 
must be included in any realistic simulation of QCD,
at present the Lattice is not near to the continuum limit
with light, dynamical quarks.
What it can do, with {\it precision} near the continuum limit, is
tell us about the pure glue theory.  While it is easy to bemoan
the lack of nearly continuum data with quarks, it should not
obscure this fundamental advance.  That we can compute reliably
the thermodynamics for non-abelian fields with three colors is
no mean feat.

In the pure gauge theory, the deconfining phase transition is
rigorously associated with the spontaneous breaking of a global
$Z(3)$ symmetry, above a transition temperature $T_c$.  Taking the
string tension to be ($400$~MeV)$^2$, 
$T_c\approx 270$~MeV, within errors which are, at present, $\approx \pm 5\%$.
One expects a confined phase of hadrons below $T_c$, and a deconfined
Quark-Gluon Plasma above $T_c$.  The great surprise from the
Lattice is that there is a third region, sandwiched between
these two.  This happens because the deconfining transition
is so weakly first order that there is a nearly critical
region about $T_c$.  For example, just
below $T_c$, the string tension is about one-tenth its value at
zero temperature.  The nearly critical modes which become light
about $T_c$ are electric $Z(3)$ glueballs, or Polyakov loops.
Magnetic $Z(3)$ glueballs, or 't Hooft loops, stay heavy.

The Lattice also gives us insight into
the transition as the number of colors is varied from three.
It is truly second order for two colors, with a 
critical region whose width, in terms of the reduced temperature
$T/T_c$, is wider than for three colors.
This leads me to the conjecture that for $N_c$ colors,
while the transition is of second order when $N_c < \infty$,
its width, in terms of $T/T_c$,
shrinks like $1/N_c$ as $N_c \rightarrow \infty$.  At $N_c = \infty$,
then, the transition is of first order.

For three colors, the nearly critical region is within
$\approx \pm 10\%$ of $T_c$.
A consequence of a narrow critical region
is that the potentials for Polyakov loops change {\it very} rapidly,
within temperatures $\approx \pm 2.5\%$ of $T_c$ \cite{dumitru}.
For the rapid expansion rates present in heavy ion collisions, this
suggests that even if the system is thermal at $T_c$, its
evolution below $T_c$ is far out of equilibrium \cite{dumitru}.  

A recent advance on the Lattice is to compute
spectral functions using
the Maximum Entropy Method \cite{karsch}.  The spectral density
for the pion has a narrow peak
in the confined phase, but is a broad peak
in the deconfined phase, at least by $1.5 T_c$.  
In the $\rho$ channel, at $.6 T_c$ the peak has moved up a little,
and broadened more;
the $\rho$ is completely washed out by $1.5 T_c$.
Spectral functions for heavy quark bound states have also
been computed.  Notably, the Lattice finds that the $J/\Psi$ remains bound
even at $1.5 T_c$, although less tightly bound states, such
as the $\Psi'$, evaporate below $T_c$.  This is in contrast to
potential models, which find, due to the decreasing string tension,
that all charmonium bound states disassociate
below $T_c$ \cite{wong}.

At present, the Lattice can only simulate 
dynamical quarks if the pions are too heavy.
Estimates are $T_c \approx 175$~MeV, with no true phase
transition in the thermodynamic limit, but only cross over behavior.
It is possible that a first order transition reappears for physical
pions, closer to the continuum limit \cite{kanaya}.  In
principle, the transition
can be more strongly first order than in the pure glue theory,
since with the addition of three flavors of
massless quarks, the ideal gas pressure goes up by a factor of three.
(It is useful to compare to the ideal gas pressure, since 
asymptotic freedom implies that the pressure is ideal at infinite
temperature.)

Present simulations with dynamical quarks demonstrate an
approximate universality, termed ``flavor independence''.  This
is the observation that as a function of $T/T_c$,
the ratio of the true pressure, to the corresponding
value in an ideal gas of quarks and gluons, is nearly universal,
independent of the number of flavors.  
This is remarkable, given the large changes in both $T_c$
and in the ideal gas term. This suggests
that the thermodynamics of QCD, with three colors
and three flavors of dynamical
quarks, is dominated by that of the pure glue theory.

Advances have been made with systems at nonzero (quark)
chemical potential $\mu$.  
The Lattice can generate a thermal distribution of quarks, but
with standard Monte Carlo techniques, it cannot fill a Fermi sea.
However, for a quark with energy $E$, if the temperature $T$ is
much greater than $\mu$, then 
the Fermi-Dirac distribution function 
at $\mu \neq 0$ is approximately that for $\mu = 0$:
\beq
\frac{1}{e^{(E-\mu)/T} + 1} \; \approx \;
\frac{1}{e^{E/T} + 1} 
\eeq
That is, at high temperature it doesn't matter much whether or not you fill
the Fermi sea.  Thus the Lattice can compute at $T \gg \mu$:
by $\mu \sim 200$~MeV, $T_c$ has only decreased
a small amount, to $\sim 160$~MeV \cite{fodor}.  This is
well on the way to nuclear matter, which at zero temperature,
occurs when $\mu$ exceeds $\approx 313$~MeV.
This is a new and strong constraint on effective models of thermal QCD,
which often give too large a decrease.  

\section{STATISTICAL MODELS: A BIG BOOST (VELOCITY)}

For the most central collisions at zero rapidity, an amazing
summary of the single particle spectra is a thermal fit 
\cite{ulrich,bialas,rafelski,koch}.  
Fits in which chemical freeze-out occurs at
the same temperature as kinetic freeze-out are favored,
with $T \approx 165$~MeV and $\mu \approx 14$~MeV 
\cite{florkowski}.   Resonance decays
describe the excess of low momentum pions in central collisions.
What happens for peripheral collisions, which must also
have resonance decays, and yet do not have an excess of
low momentum pions, is less clear, and usually ignored.

The approximate equality of the temperatures for chemical and
kinetic freeze-out is peculiar.  Any scattering in a hadronic
phase produces chemical freeze-out at a higher temperature than
that for kinetic freeze-out \cite{bleicher,greiner}, 
so the data suggest that both
temperatures are really one of hadronization, with little rescattering
in a hadronic phase.  This is one hint of possible non-equilibrium
behavior at RHIC.

The temperature for chemical freeze-out is consistent with data at lower
$\sqrt{s}/A$.  From energies of $\sqrt{s}/A$ from a few GeV on up,
chemical freeze-out occurs along a curve in which
the energy per particle is constant, about $\approx 1$~GeV.
In the plane of $T$ and $\mu$, even if the hadronization temperature
agrees with $T_c$ at $\mu =0$, it
is distinctly lower than $T_c(\mu)$ for $\mu \neq 0$.
For example, chemical freeze-out at 
AGS energies gives about $\mu = 200$~MeV,
and a hadronization temperature which is at most $\approx 120$~MeV;
from the Lattice, though, at this $\mu$ the corresponding
$T_c$ is much higher, $\approx 160$~MeV \cite{fodor}.

To describe the behavior of particles with increasing mass,
it is necessary to assume that {\it all} hadrons are emitted with respect
to a local moving rest frame.  At RHIC, the radial velocities of
this local rest frame go up to
$\approx 2/3$~c; averaged over radius,
they are about $\approx 1/2$~c.  This can be seen by eye: versus
$p_t$, single particle distributions for pions turn up, while those
for protons (say) turn down.

The radial dependence of the velocity of the local rest frame is
not constrained by the data, and is fit to agree with the observed
spectrum.  The same is true of hydrodynamical models
\cite{huovinen}.  This is why they are fits.
For example, consider how the single particle
spectra change with rapidity, or centrality.  While the temperature
{\it might} be the same, the local flow velocity now depends 
not just upon the radius, but also upon the
rapidity, centrality, {\it etc.}  It is untenable to
consider only zero rapidity, and ignore the rest.

A statistical model implies not only what the chemical composition
is, but, as well, the $p_t$-dependence of the single particle
spectrum.  Of course a thermal distribution should only hold up
to some upper scale, perhaps $1-2$~GeV.  It would be interesting
to compute the ratios of moments of transverse momenta:
\beq
r_n \; = \; 
\frac{|\langle p_t^n\rangle_{exp} - \langle p_t^n\rangle_{th}|}
{\langle p_t^n\rangle_{th}} \; .
\eeq
Here $exp$ and $th$ denote, respectively, 
moments computed from experiment, versus
a thermal distribution (with some assumed velocity profile).
By definition, if the overall number of particles is thermal,
$r_0=0$.  For $n>1$, $r_n$ is a dimensionless series of pure numbers;
the fit is good until $r_n$ is no longer small.  This must
happen at some large $n$, since eventually fluctuations from
hard momentum processes dominate.  It would
be interesting to determine these ratios from experiment, for
all collisions in which a thermal fit works.

\section{``IDEAL'' HYDRODYNAMICS AND ELLIPTICAL FLOW}

A dynamical realization of a thermal fit is a hydrodynamical model.
A specific equation of state is assumed, with some initial temperature
and velocity profile at some initial time.  The system then evolves
according to an equation of state, with the parameters chosen
to agree with the observed single particle spectra.
First order transitions are usually assumed, but 
this doesn't matter much, given
the rapid expansion characteristic of heavy ion collisions.

A measure of hydrodynamic behavior is given by elliptic flow.  For
a peripheral collision, in which the two nuclei only partially overlap,
an ``almond'' is formed in the plane perpendicular to the reaction
plane.  As the system
hadronizes, this spatial anistropy turns into a momentum anistropy,
with the average momentum larger along the narrow part of the almond
then along the long part.  This elliptical anistropy \cite{huovinen,pratt}
has been measured as a function of centrality and $p_t$; overall, the
values at RHIC are about twice as large as at the SPS.  By
geometry, elliptic flow vanishes for zero centrality, as nuclei
which completely overlap cannot have any anistropy.  
Elliptic flow, as measured from two-particle correlations,
fails to do this, so that more sophisticated measures, involving
correlations between four or more particles, are imperative \cite{borghini}.  

Hydrodynamic models predict that for pions, the elliptic flow
depends linearly on the transverse momentum.  The local flow
velocity also predicts the behavior of elliptic flow for heavier
particles, such as protons.  Both predictions are borne out by
the experimental data, for momenta up to $p_t \approx 1.5$~GeV.

Versus centrality, as measured by the number of participants,
hydrodynamics predicts that 
the elliptic anistropy is linear near zero centrality, which is
observed.  When the number of participants is half the maximum
value, though, hydrodynamics significantly overpredicts the
elliptic flow.

Calculations with three dimensional
hydrodynamics, giving single particle distributions versus
rapidity, have been performed \cite{hirano}.
Even with agreement at zero rapidity, the results from these
hydrodynamic calculations are much broader in rapidity than
the experimental data \cite{voloshin,huovinen}.

There are two conundrums about these hydrodynamic fits:
first, the initial times required in hydrodynamic calculations are 
{\it extremely} small, $\approx .6$~fm/c \cite{heinz}.  
Secondly, they assume ideal hydrodynamics.
Even with times expected from saturation (see below), this
is a very short time.  Further, in QCD
viscosity coefficients can be computed from the Boltzman-Altarelli-Parisi
equations.  At least in weak coupling, 
the shear viscosity is large (on a natural scale),
and cannot be neglected \cite{huovinen,teaney}.  

These technicalities should not obscure the fact that elliptic flow
demonstrates that heavy ion collisions exhibit significant collective
behavior.  
Further, while fits to single particle spectra with
ideal hydrodynamics and short times may well work, this does not exclude
the possibility of similar fits with non-ideal hydrodynamics;
it is not clear how the times might then change.

As an example, \cite{heinz} assumes that particles obey a thermal
distribution in the transverse direction, but free stream in the
longitudinal (or beam) direction.  
Adjusting the initial conditions to give the right
single particle distributions, the resulting elliptical anistropy
is too small by a factor of two.  As in many other examples, the
data sharply limits theory.

Experimentally, it is unremarkable 
that hydrodynamics fails above $p_t \approx 1.5$~GeV.
Hydrodynamics should break down at short distances;
that it works down to $\approx .13$~fm is actually pretty good.
Rather, the surprise is that the elliptic anistropy is approximately 
{\it constant} for $p_t: 2 \rightarrow 6$~GeV.  In QCD, one expects
cross sections to peak at some momentum scale on the order of
a few GeV, and then to fall off with the powers 
characteristic of QCD.  It is very difficult to imagine how anything
flat in $p_t$ could ever emerge.  

One possibility is that one is not measuring collective phenomenon
at all, but simply some property of two jets, which is misinterpreted
as collective flow.  
\cite{kovchegov} used a mini-jet model to obtain
an elliptic anistropy which agrees with experiment; the single-particle
spectra, though, is wrong.
Methods to determine elliptic anistropy using correlations between
four or more particles should be insensitive to 
contamination by jets \cite{borghini}.  

\section{HBT RADII: A ``BLAST'' WAVE?}

For identical particles, a length scale can be determined by
pion interferometry through the Hanbury-Brown-Twiss (HBT) effect
\cite{pratt}.  This length scale is related to the surface at
which the pions last interacted.
Since there is axial symmetry to a heavy ion collision, there are
three distances, corresponding to along the beam direction, $R_{long}$,
along the line of sight, $R_{out}$, and perpendicular to that,
$R_{side}$.  

One of the big surprises from RHIC is that the HBT radii did {\it not}
grow much between $\sqrt{s}/A = 17$ to $200$~GeV.  The change
in $\rlong \rside \rout$ is, more or less, the same as the increase in
multiplicity, $\approx 50\%$.  

This can be taken as direct experimental evidence for the {\it absence}
of a strongly first order phase transition in QCD, completely 
independent from the Lattice.  If the transition were strongly
first order, as it went through $T_c$ the system would supercool
and grow in size.  Estimates of the sizes of the system before QM'01
ranged up to tens of fermi, which are not seen.  
Unfortunately, putting a bound on the
latent heat of the transition is manifestly a model dependent
exercise.  Still, it would be an amusing exercise.

The details of the HBT radii, however, have proven to be {\it much}
more interesting than expected.  
Before the RHIC data, it was thought that the hadronic firetube
from an $AA$ collision might be like a ``burning log''.  But
instead of smouldering, the RHIC data suggests that the log blows up.

In particular, the results from RHIC 
appear to contradict {\it any} hydrodynamic description \cite{pratt,greiner}.
Versus experiment, hydrodynamics gives values of 
$\rlong$ and $\rout$ which are too large, and a
$\rside$ which is too small.  Of especial interst is the ratio
of $\rout/\rside$: hydrodynamics predicts this ratio should be
$\approx 1.5 \rightarrow 2$, and which increases with $p_t$.
At RHIC, the ratio decreases as $p_t$ goes up, and is
about one, $.85 \geq \rout/\rside \geq 1.15$ \cite{pratt}.

The HBT data can be parametrized as a type of ``blast'' wave,
with a velocity $\approx 3/4$~c \cite{pratt}.  This may
indicate a type of ``explosive'' behavior \cite{dumitru,shuryak},
a term first used by the experimentalists.

HBT radii can also be used to give a measure of the entropy
carried by pions \cite{pratt}.  
Per degree of freedom, the entropy in
peripheral collisions is greater than that in a Bose-Einstein
gas, but is less than that for the most central collisions.
This may indicate coherence in the initial state.

Partonic models in which the cross section is artificially multiplied
by a factor of $\approx 14$ \cite{molnar} give both
$\rout/\rside \approx 1$, at least by $p_t \sim 1.5$~GeV, and a flat
elliptical anistropy.  
One may view this as evidence that at hadronization, 
apparently near $T_c$, the partons appear
to behave {\it much} more strongly than expected from
perturbative QCD.

\section{``HIGH''-$p_t$ SUPPRESSION}

From the first RHIC data, it was clear that
the spectra for ``high''-$p_t$ particles, 
meaning above, say, $2$~GeV, is qualitatively
different in central $AA$ collisions, versus $pp$ collisions at the
same energy.  Dividing by the number of participants,
the number of particles at high-$p_t$ is significantly 
less in central $AA$ collisions than in $pp$, 
by overall factors of $2-4$ \cite{kunde,miod,baier,muller,wang}.
This is quantified through the ratio $R_{AA}$, which is the
ratio of the number of particles in central $AA$ collisions, divided
by that in $pp$, as a function of $p_t$.  The suppression begins
above $p_t \approx 2$~GeV; above $4$~GeV, $R_{AA} \approx 1/3\rightarrow
1/4$ for charged hadrons, and 
$R_{AA} \approx 1/5 \rightarrow 1/6$ for pions \cite{miod}.  A surprise
of the Run II data is that for
$p_t: 2 \rightarrow 9$~GeV, $R_{AA}$ is approximately constant,
up to at least $9$~GeV \cite{miod}.

This suppression of high-$p_t$ particles is opposite to what is
observed at the SPS.  There, due to what is known as the Cronin
effect, the ratio $R_{AA}$ is greater than one,
going up to $\approx 2.5$ by $p_t \approx 3$~GeV.
This change in the spectrum must be considered as one of the
most dramatic features of the RHIC data.

The usual explanation of high-$p_t$ suppression is due to
energy loss \cite{baier,muller,wang}.  Bjorken originally
noted that a fast quark (or gluon) loses energy as it traverses
a thermal bath, in just that same way that any charged particle
does in matter.  Single particle distributions
can be explained using parton models \cite{wang}.

Detailed features of the data appear difficult to
explain as energy loss, though.  Naively, 
one might expect soft processes to scale
as the number of participants, $\sim A$, and hard processes
to scale as the number of collisions, $\sim A^{4/3}$.
As a hard process, then, one would expect energy loss to scale
as $A^{4/3}$.  Instead, in peripheral collisions,
$R_{AA}$ scales with
the number of participants \cite{baker}.  It is possible that
the difference of two terms of order $A^{4/3}$ is of order
$A$, but that seems unnatural.

Further, the observed constancy of $R_{AA}$ for $p_t: 2 \rightarrow
9$~GeV is surprising; perturbative models of QCD do not give constant
behavior.  The apparent constancy also reflects changes in particle
composition, while pions dominate below $p_t \approx 2$~GeV, 
unlike $pp$ collisions, there are as many protons as pions above
$p_t \approx 2$~GeV.  Taking
protons as color baryon junctions, in perturbative QCD
the pion part of $R_{AA}$ peaks at a smaller $p_t$
than the proton part \cite{vitev}.  

In weak coupling, energy loss in QCD is proportional to $\sqrt{E}$,
where $E$ is the energy of the jet \cite{baier}.  If one assumes,
simply for the sake of understanding the data, that the energy
loss is proportional instead to $E$, with hard particles
losing $\approx 7\%$ of their energy per scattering, 
then a next-to-leading order calculation in QCD gives
a good fit to the data \cite{sarcevic}.  

\section{SATURATION}

Another surprise from the first RHIC data was that the multiplicity
did not grow as rapidly as predicted, at least 
on the basis of various cascade
models.  A natural explanation for this is given by models of 
saturation \cite{muller,kharzeev,iancu,krasnitz,ruuskanen,tuominen}.

The application of saturation to AA collisions is, at the most basic
level, purely a kinematical effect.   Consider a nucleus-nucleus 
collision, in the rest frame of one of the nuclei.  
For atomic number $A \approx 200$,
in its rest frame the incident nucleus has a diameter no greater than 
$\approx 2 A^{1/3} \approx 15$~fm.
By Lorentz contraction, this distance gets shrunk down by
a factor which is about $1/(\sqrt{s}/A)$.  
Eventually, the color charge of the incident nucleus looks not like
a nucleus, but just like a very thin pancake, with a big color
charge $\sim A^{1/3}$.  
Assuming that distances on the order
of $1/3 \rightarrow 1/4$~fm 
are small on hadronic scales, the incident nucleus looks like a thin
pancake when $\sqrt{s}/A: 45 \rightarrow 60$~GeV.
It is amusing that a simple
estimate gives an energy right near where a Central Plateau, in 
which the particle density is constant with rapidity, first appears.

In detail, saturation is a dynamic criterion.  It states that
at sufficiently small Bjorken-$x$, quark and gluon distribution
functions are dominated by gluons, which peak at a characteristic
momentum scale, termed the ``saturation'' momentum, $p_{sat}$.
(This gluon dominance is reminiscent of flavor independence
for thermodynamics.)
For any perturbative approach to work, $p_{sat}$ cannot
be less than at least $1$~GeV.  The above kinematic argument
suggests that $p^2_{sat} \sim A^{1/3}$: thus one can probe smaller
$x$ values with large nuclei at RHIC, say, than in $ep$ collisions
at HERA.

What is most important about saturation is, again,
almost a kinematical effect: it resets the ``clock''
for heavy ion collisions.  In the Bjorken picture which dominated before
the RHIC data, one assumed that hadronization occured at time scales
$\approx 1$~fm/c; after all, what other time scale is there?  Thus
in the Bjorken picture, there seemed as if there
was little time for even the largest nuclei, 
only $7$~fm in radius, to thermalize.  (Unless, again, there were
a strongly first order transition, which is why it was so popular before
RHIC.)

With saturation, however,
the natural scale of the clock is given by $1/p_{sat}$;
for $p_{sat} \approx 1$~GeV, this is already $\approx .2$~fm/c;
see, {\it e.g.}, \cite{ruuskanen}.  That
is, saturation makes the hadronic ``clock'' runs at least {\it five} times
faster!  The possibility of interesting things happening is far more likely.  

Saturation is realized in the 
Color Glass Model \cite{muller,kharzeev,iancu}.  The gluon fields
from the incident nucleus are described as classical color sources,
reacting much quicker than the fields in the target nucleus.  Taking
a gluon field to scale with the QCD coupling constant $g$ as
$A_\mu^a \sim 1/g$, one concludes that the action, and indeed
all quantities --- such as particle multiplicity, average energy,
{\it etc.} --- scale like $1/g^2$.  In
an asymptotically free regime, then, all quantities grow like
$\sim 1/\alpha_s(p_{sat}) \approx \log(p_{sat})$.  This small,
logarithmic growth in the multiplicity agrees qualitatively with
the RHIC data (although one really needs the increase to LHC energies
to make this quantitative).

This picture is only approximate.  Even if a gauge field is
$\approx 1/g$, the action need not scale like $1/g^2$.
In $AA$ collisions, at initial stages there is a screening
mass generated along the beam direction, but not transverse, with a mass
squared $\sim \alpha_s$ at leading order \cite{baier,muller}.
Such a dynamically generated mass scale changes
integral powers of $1/\alpha_s \sim \log(p_{sat})$ into
fractional powers; see, {\it e.g.}, \cite{baier,muller}.

Modulo these theoretical quibbles,
it seems plausible that saturation describes the initial state of
$AA$ collisions at high energies.  Fits to the particle multiplicity,
including the dependence upon centrality and rapidity, agree
approximately with the data \cite{kharzeev}.  
It is not evident how to turn gluons into hadrons, as
sometimes the mysteries of ``parton-hadron duality'' are invoked.
It is surprising
that such models work over a wide region of
centrality and rapidity, since
saturation (valid at small Bjorken-$x$)
should not work well in the fragmentation region (which is large
Bjorken-$x$).  The particle density in such fits has
peculiar peak at zero rapidity \cite{tuominen}.
Experimentally, the fragmentation
region appears to be described by a universal limiting form \cite{baker}
(although it is crucial to have data with identified particles,
which is presently lacking).

Saturation does not describe other basic features of the data, though.
The most serious problem is that the average $p_t$ in saturation
is $\langle p_t \rangle \approx 2 p_{sat}$; even with $p_{sat}$
as low as $1$~GeV, this is an average $p_t \sim 2$~GeV.  In contrast,
at RHIC $\langle p_t \rangle \approx 550$~MeV.  The average energy
from saturation will decrease due to inelastic processes and
the generation of entropy.  Assuming that this fixes the
overall constant, one is still at a loss to explain why the average
$p_t$ changes by at {\it most} $1 \%$ between $\sqrt{s}/A = 130$~GeV
and $200$~GeV, while the multiplicity changes by at least
$15\%$ \cite{ulrich}.  In saturation models, the average $p_t$
grows with multiplicity.

A related problem is the chemical composition.  Parton-hadron duality
is really gluon-pion duality; but if the average gluon momentum is
large, why don't the hard gluons become kaons?  Instead, at RHIC
kaons are much less numerous than pions, only about $15\%$ as much.

The elliptic anisotropy can be calculated
in a Color Glass \cite{krasnitz}.  As a function of momentum, it
is dominated by very soft momentum, peaking at a value of
$p_t \approx p_{sat}/4$.  This is not seen experimentally.

Of course if saturation describes the initial state, and not the
final state, then there is no problem with the above features of
the data.  Estimates of thermalization in perturbative QCD
give numbers $\approx 3$~fm/c \cite{muller,baier}.  
Further, thermalization occurs
not due to elastic processes, but is dominated by inelastic
processes \cite{serreau}.

It is claimed that the logarithmic growth of the total cross section
follows from saturation \cite{iancu}.  This is a matter of black and
white: if there are massless particles, there is no bound on the
total cross section.  Thus saturation, which has no mass gap, cannot
guarantee a finite total cross section.


\begin{thebibliography}{9}
\bibitem{satz} H. Satz, hep-ph/0209181.
\bibitem{ramello} L. Ramello, http://alice-france.in2p3.fr/qm2002/Transparencies/23Plenary/Ramello.pdf
\bibitem{wessels} J. Wessels, http://alice-france.in2p3.fr/qm2002/Transparencies/23Plenary/Wessels.pdf
\bibitem{wambach} J. Wambach, http://alice-france.in2p3.fr/qm2002/Transparencies/24Plenary/Wambach.pdf
\bibitem{bearden}  L. Bearden, http://alice-france.in2p3.fr/qm2002/Transparencies/20Plenary/Bearden.ppt
\bibitem{ulrich} T. Ullrich, nucl-ex/0211004.
\bibitem{vanburen} G. Van Buren, nucl-ex/0211021.
\bibitem{kunde} G. Kunde, nucl-ex/0211018.
\bibitem{miod} S. Mioduszewski, nucl-ex/0210021.
\bibitem{baier} R. Baier, hep-ph/0209038.
\bibitem{muller} A. Mueller, hep-ph/0208278.
\bibitem{wang} X. N. Wang, nucl-th/0208079.
\bibitem{voloshin} S. Voloshin, nucl-ex/0210014.
\bibitem{huovinen} P. Huovinen, nucl-th/0210024.
\bibitem{pratt} S. Pratt, http://alice-france.in2p3.fr/qm2002/Transparencies/24Plenary/Pratt.ppt
\bibitem{hardtke} D. Hardtke, nucl-ex/0212004.
\bibitem{kanaya} K. Kanaya, hep-ph/0209116.
\bibitem{dumitru} A. Dumitru, nucl-th/0209001.
\bibitem{karsch} F. Karsch, hep-ph/0209028.
\bibitem{wong} C. Y. Wong, nucl-th/0209017.
\bibitem{fodor} Z. Fodor, hep-lat/0209101.
\bibitem{bialas} A. Bialas, http://alice-france.in2p3.fr/qm2002/Transparencies/19Plenary/Bialas.zip
\bibitem{rafelski} J. Rafelski, nucl-th/0209084.
\bibitem{koch} V. Koch, nucl-th/0210070.
\bibitem{florkowski}  W. Florkowski, nucl-th/0208061.
\bibitem{bleicher} M. Bleicher, http://alice-france.in2p3.fr/qm2002/Transparencies/19Plenary/Bleicher.ps
\bibitem{greiner} C. Greiner, nucl-th/0209021.
\bibitem{borghini} N. Borghini, nucl-th/0208014.
\bibitem{hirano} T. Hirano, nucl-th/0208068.
\bibitem{heinz} U. Heinz, nucl-th/0209027.
\bibitem{teaney} D. Teaney, nucl-th/0209024.
\bibitem{kovchegov} Y. Kovchegov, Proc.
\bibitem{shuryak} E. Shuyrak, http://alice-france.in2p3.fr/qm2002/Transparencies/23Plenary/Shuryak.zip
\bibitem{molnar} D. Molnar, Proc.
\bibitem{baker} M. Baker, http://alice-france.in2p3.fr/qm2002/Transparencies/19Plenary/Baker$\underline{\;\;}$QM2002.ppt
\bibitem{vitev} I. Vitev,  hep-ph/0208108.
\bibitem{sarcevic} I. Sarcevic, nucl-th/0211084.
\bibitem{kharzeev} D. Kharzeev, nucl-th/0211083.
\bibitem{iancu} E. Iancu, hep-ph/0210236.
\bibitem{krasnitz} A. Krasnitz, hep-ph/0209341.
\bibitem{ruuskanen} P. V. Ruuskanen, nucl-th/0210005.
\bibitem{tuominen} K. Tuominen,  hep-ph/0209102.
\bibitem{serreau} J. Serreau, hep-ph/0209067.
\end{thebibliography}
\end{document}